**Identifying the Factors that Influence Urban Public Transit Demand**

CEE 7420: Urban Public Transportation

Submitted to:

Daniel A. Badoe, Ph.D.

Submitted by:

Armstrong Aboah

Lydia Johnson

Setul Shah

December 3, 2019

# EXECUTIVE SUMMARY

The rise in urbanization throughout the United States (US) in recent years has required urban planners and transportation engineers to have greater consideration for the transportation services available to residents of a metropolitan region. This compels transportation authorities to provide better and more reliable modes of public transit through improved technologies and increased service quality. These improvements can be achieved by identifying and understanding the factors that influence urban public transit demand. Common factors that can influence urban public transit demand can be internal and/or external factors. Internal factors include policy measures such as transit fares, service headways, and travel times. External factors can include geographic, socioeconomic, and highway facility characteristics.

There is inherent simultaneity between transit supply and demand, thus a two-stage least squares (2SLS) regression modeling procedure should be conducted to forecast urban transit supply and demand. As such, two multiple linear regression models should be developed: one to predict transit supply and a second to predict transit demand.

It was found that service area density, total average cost per trip, and the average number of vehicles operated in maximum service can be used to forecast transit supply, expressed as vehicle revenue hours. Furthermore, estimated vehicle revenue hours and total average fares per trip can be used to forecast transit demand, expressed as unlinked passenger trips. Additional data such as socioeconomic information of the surrounding areas for each transit agency and travel time information of the various transit systems would be useful to improve upon the models developed.

## Table of Contents

## LIST OF TABLES

## LIST OF FIGURES

# INTRODUCTION

The rise in urbanization throughout the United States (US) in recent years has required urban planners and transportation engineers to have greater consideration for the transportation services available to residents of a metropolitan region. When first evaluating the market share of metropolitan travel, travel via public transit appears to steadily decline; however, the number of individuals utilizing public transport services is steadily increasing (Hughes-Cromwick and Dickens, 2019). Furthermore, there are growing concerns regarding environmental impacts and traffic congestion in urban areas.

These factors compel transportation authorities to provide better and more reliable modes of public transit through improved technologies and increased service quality. These improvements can be achieved by identifying and understanding the factors that influence urban public transit demand. This can be done by analyzing cross-sectional travel behavior surveys, transit databases, and passenger counts on transit systems. Common factors that can influence urban public transit demand can be internal and/or external factors. Internal factors include policy measures such as transit fares, service headways, and travel times. External factors can include geographic, socioeconomic, and highway facility characteristics.

The objective of the project is to develop an understanding of the variables that influence transit patronage in metropolitan regions. This will be done by analyzing National Transit Database (NTD) files, which are comprised of transit system travel information collected from various agencies across the US. Cross tabulations, charts, and development of trial models will be generated before identification of a specification that best describes the aggregates of travel choices in metropolitan regions reported in the database can be ascertained.

The contents of this report are organized as follows: (1) a review of relevant literature regarding the investigation and determination of factors that influence transit demand and how this demand can be modeled; (2) a descriptive analysis of the NTD files; (3) a description of the methodology that will be used to identify the factors that influence transit demand in metropolitan regions and how these factors will be used to model urban transit demand; (4) a presentation and discussion of the final model specification; and (5) concluding statements regarding the resulting outcomes of the project.

## LITERATURE REVIEW

A study conducted by Ko, Kim, S.M.ASCE, and Etezady (2019) presented system-level macroscopic analyses that identify factors that affect bus rapid transit (BRT) ridership. The study utilized Global BRT Database for cities' BRT system-related information worldwide which was supplemented by the Bus Rapid Transit Information Database. The explanatory variables employed in this study included city characteristics (population, gross domestic product (GDP) per capita), BRT components (system length, number of corridors, fleet size, fare, number of stations, median bus lanes, passing lanes, real-time information systems, integrated fare-collection systems) and metro components (number of metro lines, metro length).

The descriptive analysis from the study suggested that fare levels appear to differ significantly between cities, with the BRT system in Merida, Venezuela, providing citizens with a free-of-charge service, whereas Kent, United Kingdom, provides services at a standard fare of around USD 6.9. Also, the descriptive analysis from the study revealed that the highest distribution of BRT ridership across continents was in Latin America.

The study developed two BRT ridership models with dependent variables being total daily ridership and its normalized ridership by system length. The modeling approach used was multiple linear regression. In order to develop these models, the study first identified the appropriate specifications of the models by using the Pearson correlation test. The test revealed that daily BRT ridership has a strong relationship with fleet size, system length, and the number of corridors and stations, suggesting that an.  The study considered modeling service supply variables as having a reciprocal effect on daily BRT ridership but realized that this bidirectional relationship between the dependent variable and regressors may cause a serious regression estimation problem (that is violating a key assumption of regression which states that regressors and disturbances are uncorrelated). To solve this issue, the study considered two modeling approaches: (1) excluding the service supply variables (i.e., fleet size, system length, number of corridors and stations) and (2) developing two-stage least-squares (2SLS) regression models, which have been commonly applied as a way to remedy interactions between supply- and demand-side factors.

The developed models demonstrated that service supply levels, such as fleet size and the number of BRT corridors, were critical factors for determining ridership. Including supply variables in the model helped in explaining about 80% of the variation in total ridership and 70% of the variation in normalized ridership. After controlling the effects of service levels, the models revealed that factors such as population and fare affect total daily ridership. The 2SLS approach identified that the existence of multimetro lines can increase ridership by 41% and that the operation of both integrated fare collection and real-time information systems can boost ridership by 47%.

Taylor, Miller, Iseki, and Fink (2008) sought to analyze the determinants of transit ridership across various urbanized areas within the US and to establish which factors have the greatest influence on ridership. Factors taken into consideration included those that can be controlled by transit systems, such as fare pricing and service headways and those that cannot be controlled by transit systems, such as regional geography, socioeconomic characteristics, and auto/highway system characteristics. Cross-sectional analyses of 265 urban areas within the US were conducted such that more robust and generalizable results could be generated. Data from these urban areas were retrieved from the National Transit Database (NTD) and are representative of transit systems in the year 2000.

2SLS regression models were developed to model the relationship and simultaneity between transit supply and demand according to the various factors of transit ridership mentioned previously. Vehicle revenue hours were used to measure transit supply while passenger boardings were used to measure transit demand. The first-stage model used to estimate transit supply revealed that urbanized area population and the percent of the population voting for the Democratic candidate in the 2000 presidential election were able to explain the variation in vehicle hours of service to a high degree with a coefficient of determination of 0.8216. The significance of these variables suggests that large metropolitan areas are likely to have more transit services in proportion to the size of the urban area and that Democratic-leaning areas are more likely to support the government and public expenditures on public transit services. The second-stage model used to estimate passenger boardings identified that vehicle revenue hours explains most of the variance in transit demand. Further, six factors that are exogenous to the transit system can be used to predict the patronage of urban transit systems. They include population density, whether the urban area is located in southern states, the proportion of college students, the proportion of immigrants, and the proportion of ride-captive individuals with little or no access to personal vehicles within a metropolitan area. Two policy variables endogenous to the public transit system, service frequency, and transit fares, can also be used to explain transit demand in urban areas. In total, eight explanatory variables are included in the second-stage model to predict transit demand. These eight variables are able to explain roughly 91% of the variation experienced in passenger boardings in metropolitan regions within the US. The six external variables account for most of the observed variation in ridership. The influence of the two internal transit policy variables on transit demand is smaller compared to the external factors; notwithstanding, these two factors are influential to the model as they are able to explain roughly 5% of the observed variation in ridership.

Furthermore, the elasticities generated from the study reveal values that coincide with existing transit ridership elasticities (Balcombe, Mackett, Paulley, Preston, Shires, Titheridge, Wardman, and White, 2004). A fare elasticity of -0.43, a service elasticity with respect to vehicle hours of 1.10, and a service elasticity with respect to service frequency of 0.50 were evaluated. These elasticities suggest that higher fares drive away riders while frequent service draws in more riders. However, these elasticities do not distinguish between different modes of transit or between short-term or long-term time periods.

The findings of this study reveal that external factors outside of the control of transit agencies are influential in determining transit ridership demand. However, internal variables such as service frequency and fare levels are significant contributors of transit patronage in metropolitan areas and can be used to forecast ridership in these areas.

Thompson, Brown, and Bhattacharya (2012) developed a demand model for understanding the travel behavior of the Broward County Transit (BCT) riders. BCT is the public transit agency located in Broward County, Florida. BCT serves the metropolitan area having a population of around 1.5 million. The majority of the riders on BCT are people with lower household incomes. Also, the majority of the riders transfer from one bus to another at least one time for completing their trips. This study developed a model for work trips using BCT between origins and destinations. This paper also studies the effect of transit travel time on transit ridership.

Data used for the study were from the Central Transportation Planning Program (CTPP), which is part of the United States census for the year 2000. Broward County consisted of 921 Traffic Analysis Zones (TAZs). Origin-destination travel data for 40,436 pairs were used, which is five percent of the study area population. The dependent variable used in this study is the number of work trips between origin-destination zones in Broward County. The independent variables are classified into three categories. The first category describes the travel time between each pair of zones. The second and third categories describe the characteristics of the origin and destination zone, respectively. Travel time between each origin and destination constituted of four components: (1) in-vehicle travel time, (2) transit access time, (3) transfer time, and (4) transit vehicle wait time. There is a flat transit fare structure in Broward County, therefore fare is not used as a variable in that study.

Negative binomial regression was used to develop the model for work trips using the BCT. Variables such as population, population density, auto availability per capita, and household income are used as independent variables. The goodness of fit for this model was 0.06. Land use variables such as the density of employment, mixed-use land, and walkability had no statistically significant effect on transit ridership. The coefficient estimates of variables such as the population of the zone and household income are found to have a statistically significant impact on the dependent variable.

Based on the results, zones with more jobs and smaller area size attract more transit trips compared to larger areas. Parking fees were found to have a significant impact on the dependent variable. Out of all the components of travel time, only in-vehicle travel time was found to have a significant impact on the dependent variable. It can be interpreted from the result that ridership between zones increases with lower travel time. This study helps in understanding the impact of travel time on the transit ridership between zones. However, transit fare is not flat in many of the transit agencies, therefore it is important to take transit fare into account when developing the ridership model for other transit agencies. Transit fare helps in understanding the effect of transit fare on the transit ridership.

## DESCRIPTIVE ANALYSIS

The data used for the project were retrieved from the NTD files. These files are comprised of transit system travel information for various agencies across the US. In total, 606 agencies reported transit system information to the Federal Transit Administration (FTA) over a period of seventeen years. Information such as unlinked passenger trips, vehicle revenue hours, average trip length, and average trip fare can be found within the NTD files for each transit agency. The following tables and figures provide a descriptive analysis of the NTD data.

On average, the vehicle revenue hours (VRH) have increased with time as shown in Figure 1. The average VRH peaked from 2002 to 2009 and then declined slightly from 2010 to 2011. It then finally peaked again from 2011 to 2017.

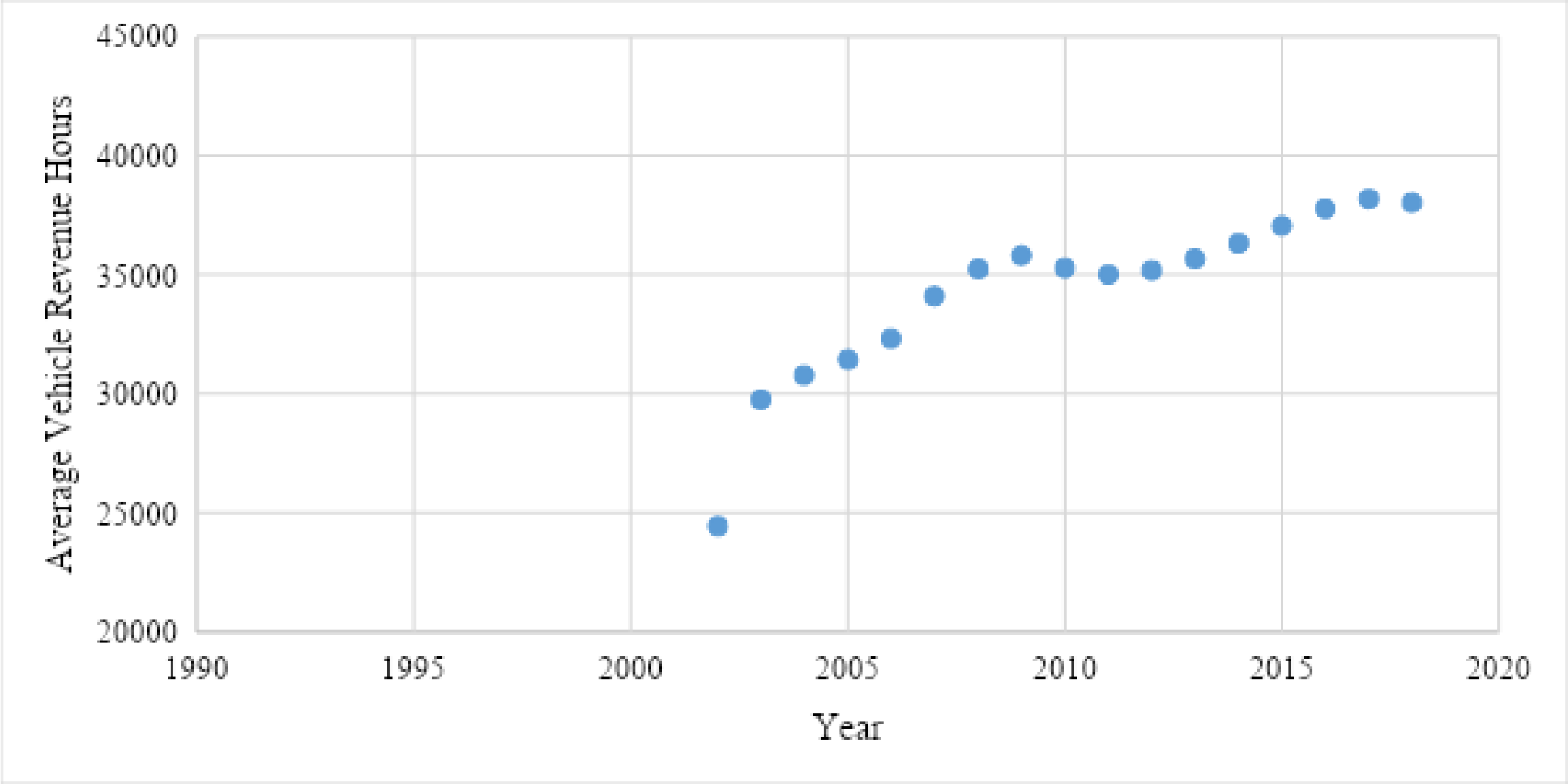

Figure 1: Average Vehicle Revenue Hours over Time

Figure 2 shows the relationship between average total unlinked passenger trips (TUPT) and time (years). The average TUPT decreased from 2002 to 2003. It then peaked steeply from 2003 to 2008 followed by a gentle decrease in 2008 to 2010. It finally peaked gently from 2010 to 2014 and then started declining gently from 2014 to 2018.

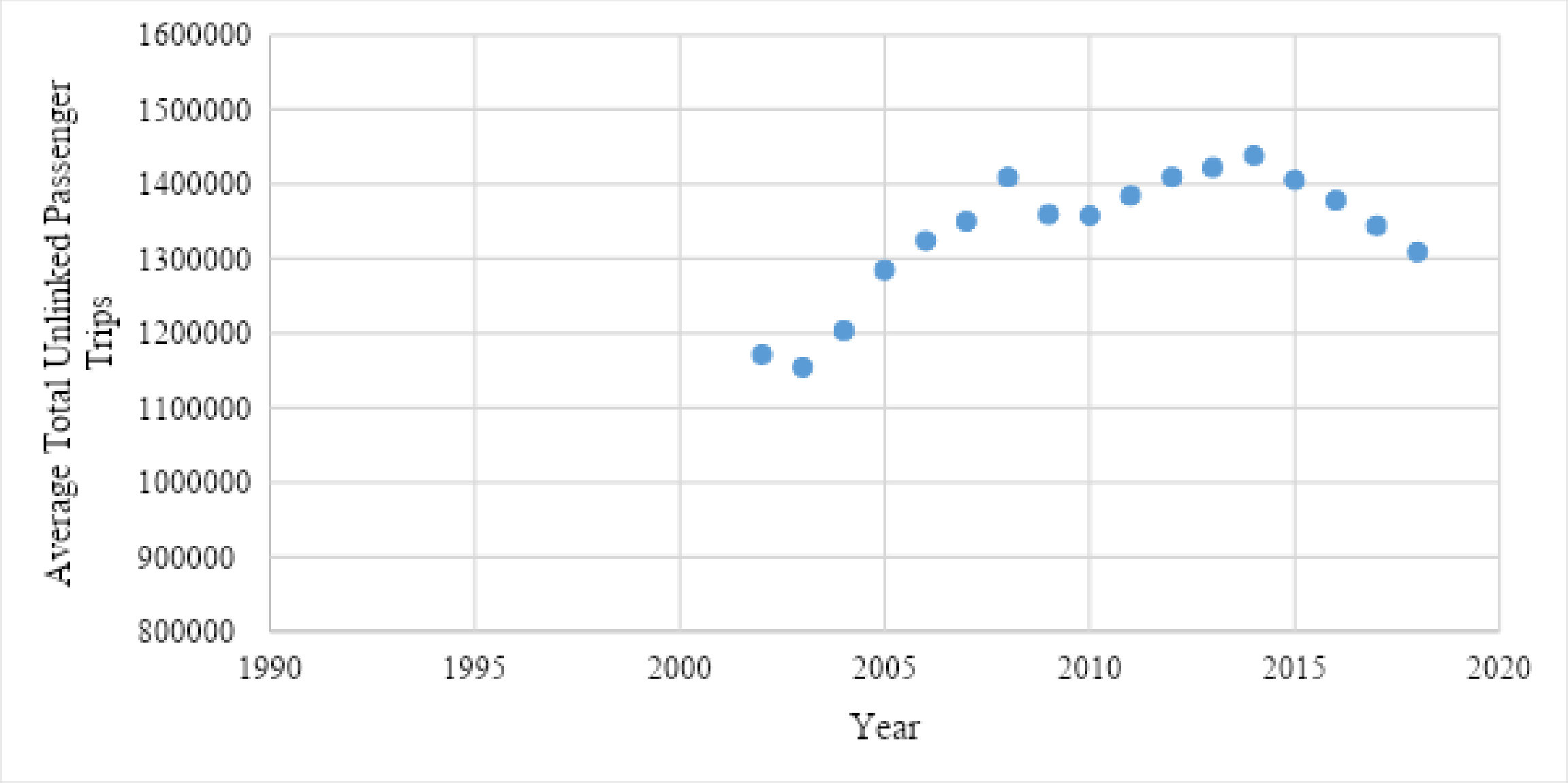

Figure 2: Average Total Unlinked Passenger Trips over Time

There is no clear relationship between the average trip length and time, as shown in Figure 3.

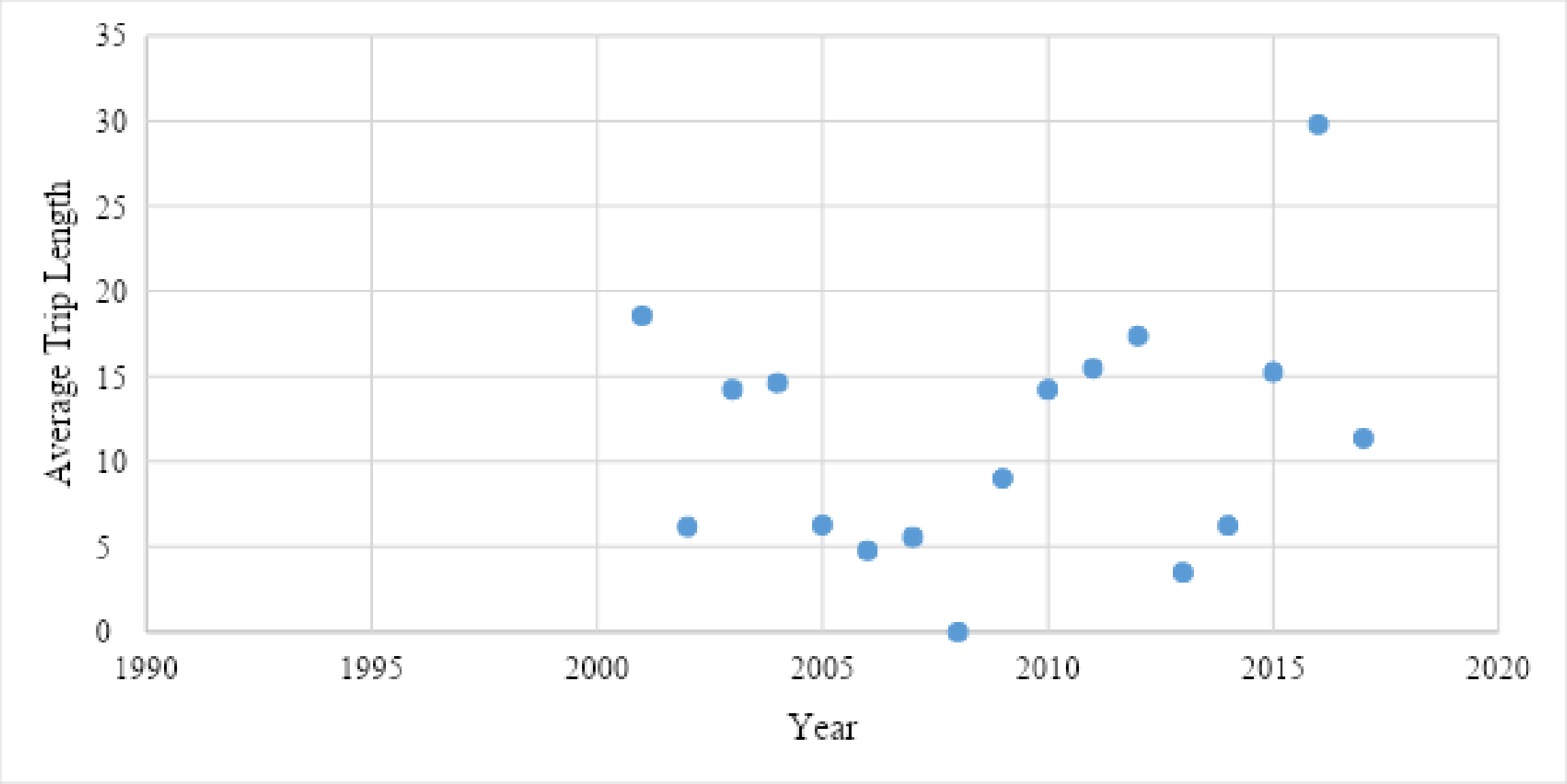

Figure 3: Average Trip Length over Time

The changes in the average number of vehicles operated in maximum service (VOMS) over time are provided in Figure 4. The figure reveals that, since 2002, the average number of VOMS has increased. This suggests that the operating expenses of transit agencies have increased over time to accommodate this increase in available vehicles. Furthermore, the trend suggests that the total number of passengers served during maximum service hours has also increased.

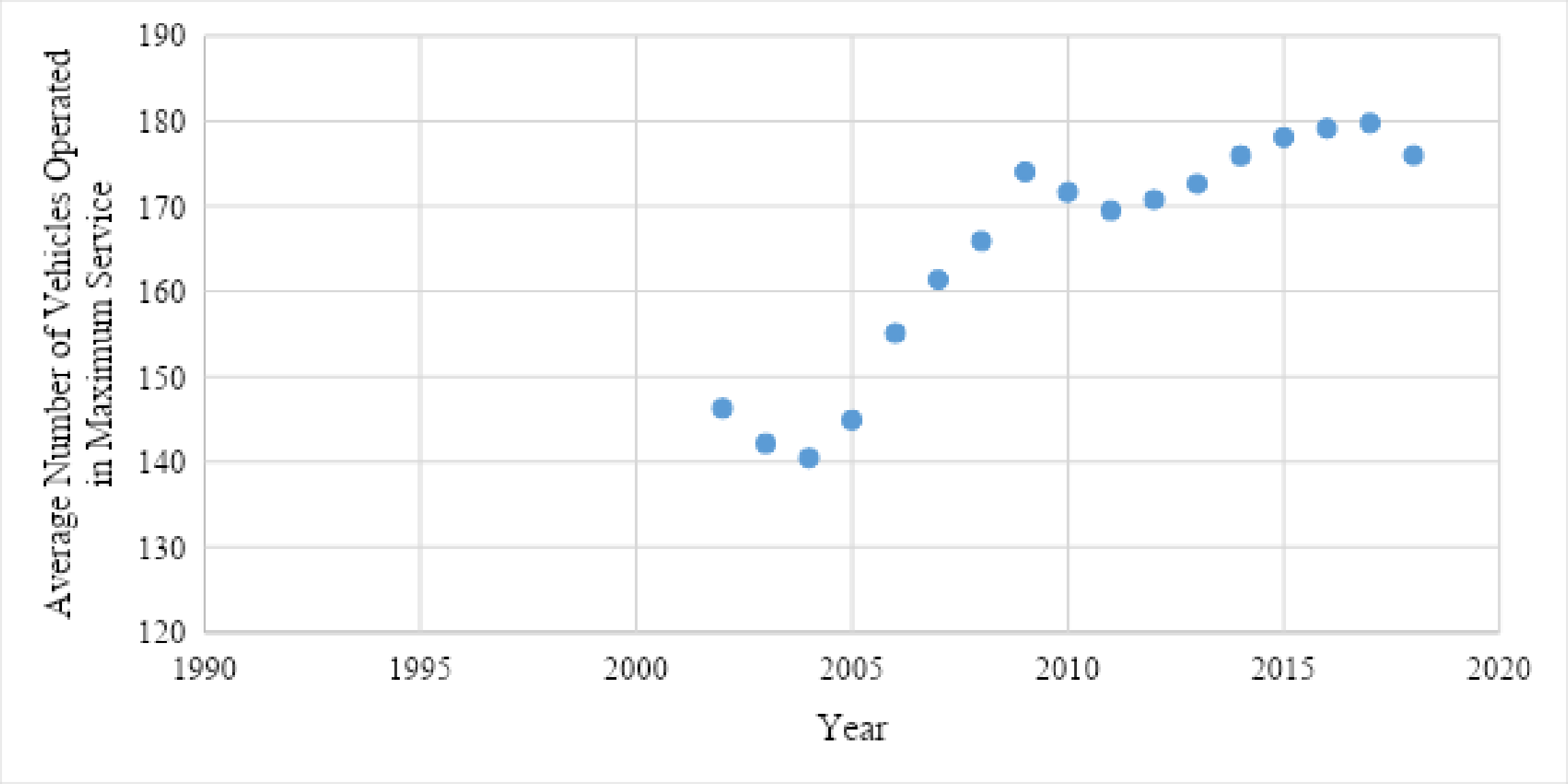

Figure 4: Average Number of Vehicles Operated in Maximum Service over Time

Figure 5 presents the average total number of passenger miles for a given year across all reporting jurisdictions. This figure reveals that with the exception of total passenger miles reported in 2017, the average miles of service generated by transit passenger demand has remained relatively stable over time.

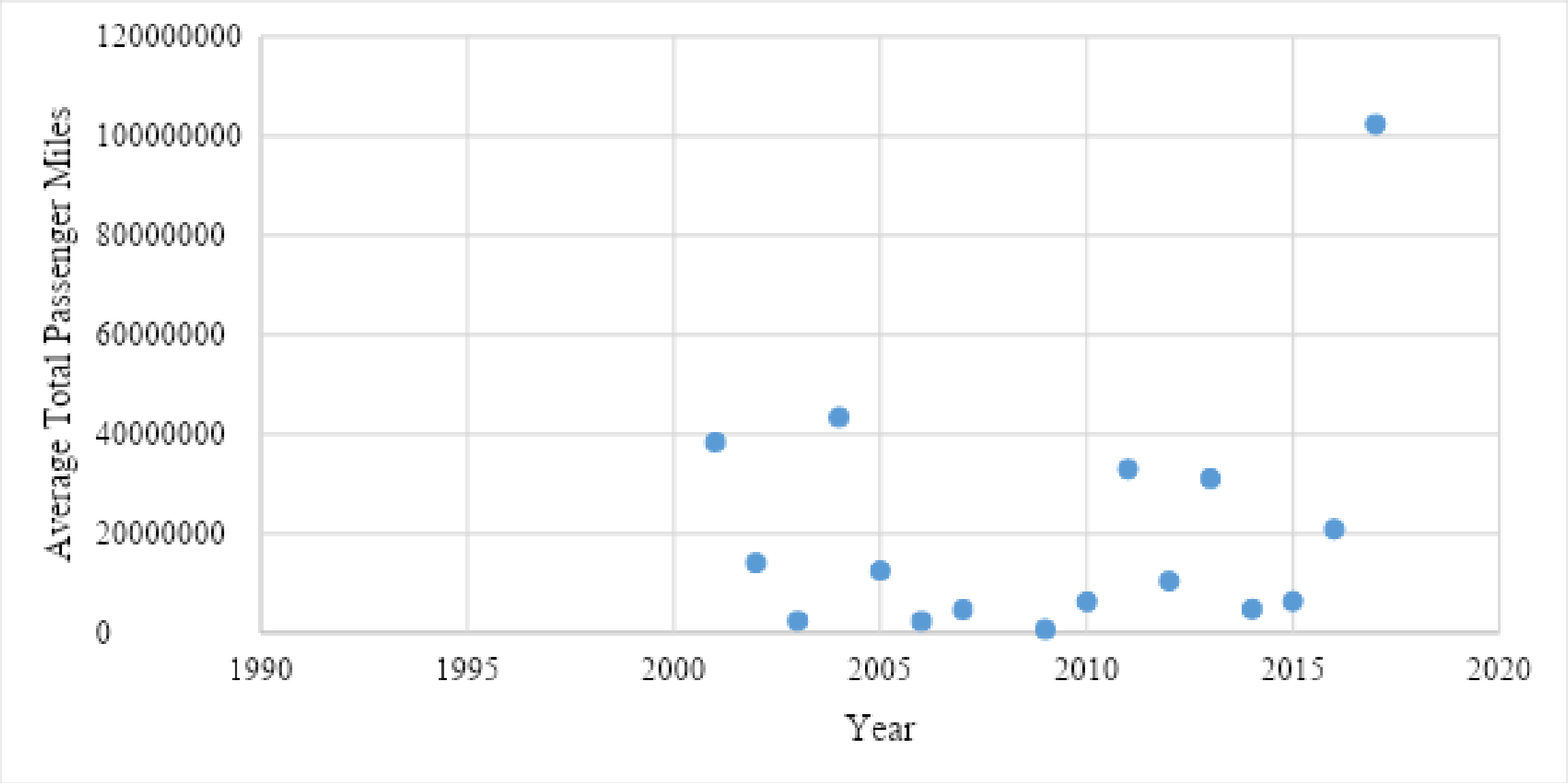

Figure 5: Average Total Passenger Miles over Time

Figure 6 presents the average service area density for a given year across all reporting jurisdictions. Service area density was generated by determining the service area population per square mileage of the service area. This figure reveals that, with the exception of service area density in 2004 and 2005, the average service area density for reporting transit agencies has remained relatively stable over time. This likely implies that the supply of urban transit has increased to accommodate an increase in service area population and passenger usage.

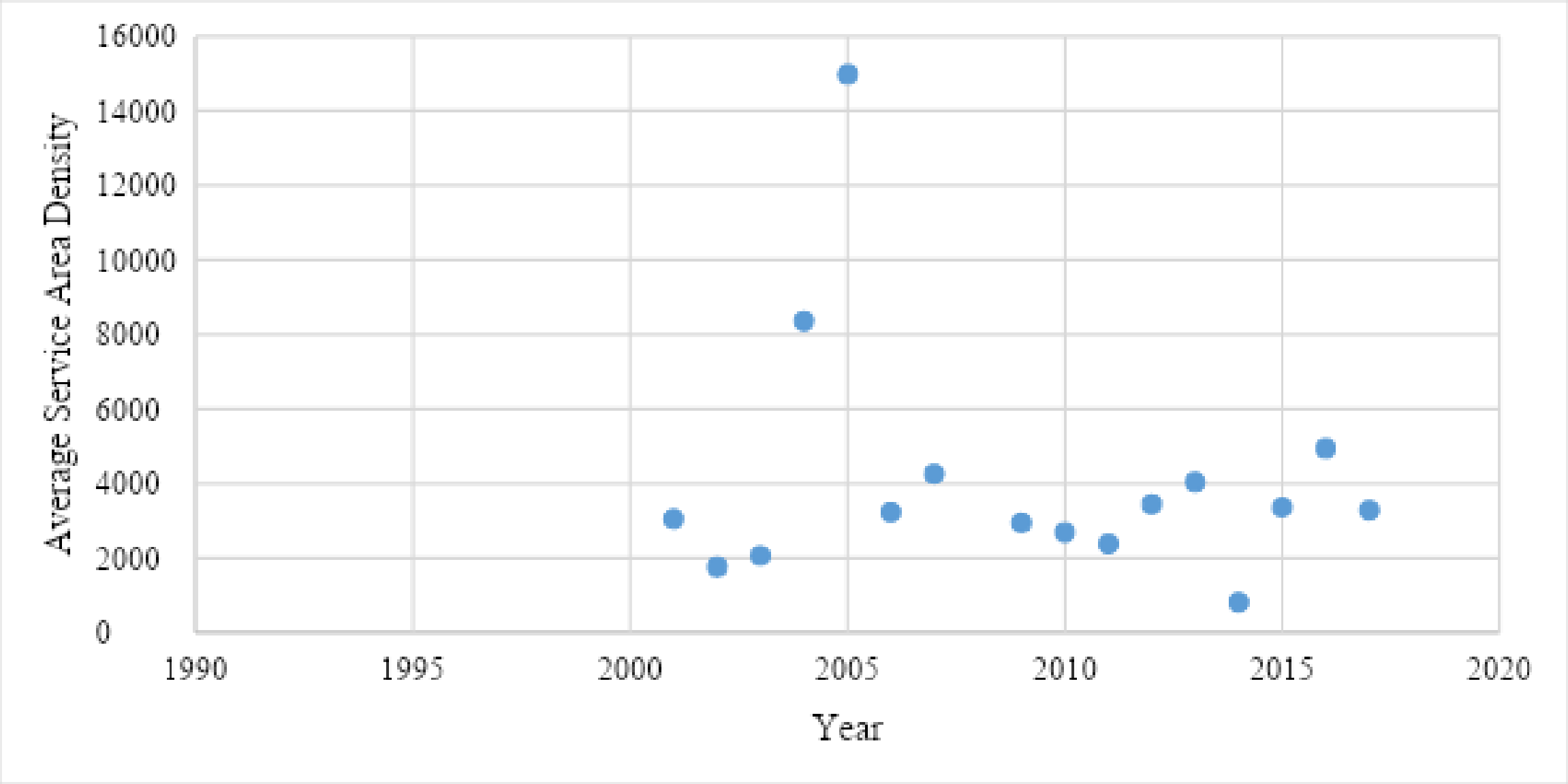

Figure 6: Average Service Area Density over Time

Figure 7 presents the average total cost per trip for a given year across all reporting jurisdictions. It can be interpreted that on average, the cost per trip has increased over time. Based on the figure below, the average total cost per trip was highest in 2009.

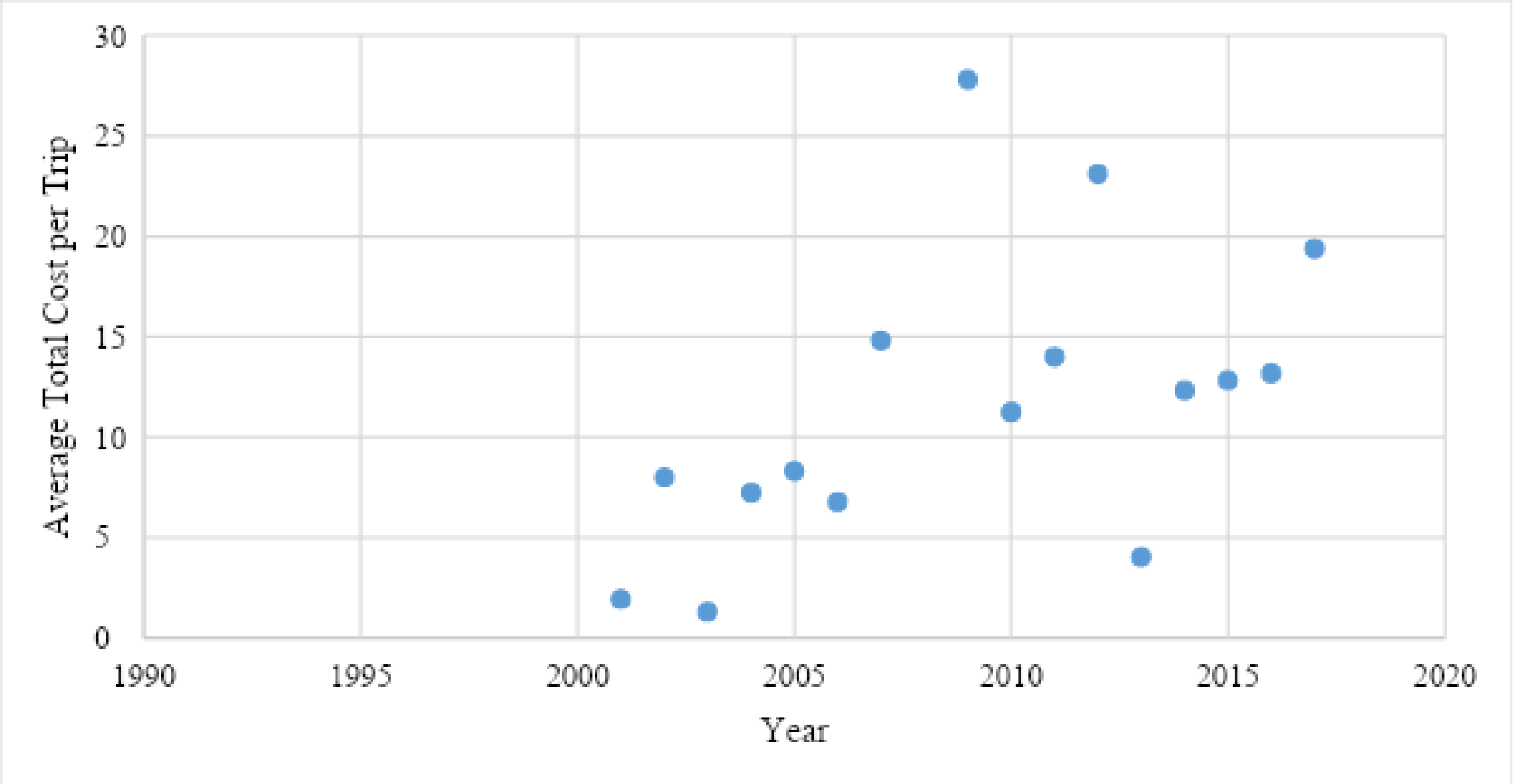

Figure 7: Average Total Cost Per Trip over Time

Figure 8 presents the average total fares per trip for a given year across all reporting jurisdictions. It can be interpreted that on average, the fare per trip has increased over time. Based on the figure below, 2012 and 2016 having a higher average fare cost per trip compared to that of other reported years.

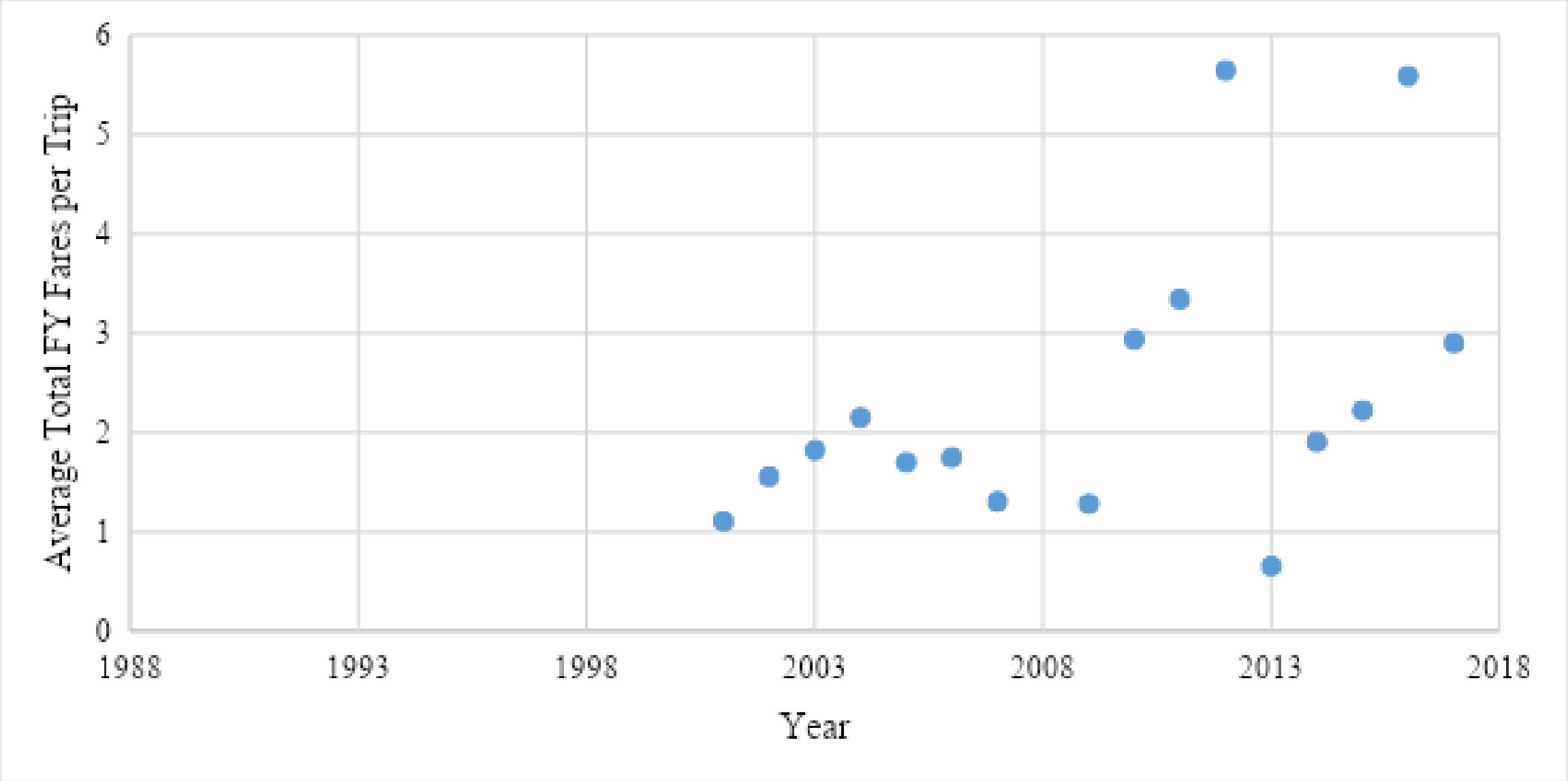

Figure 8: Average Total FY Fares per Trip

Figure 9 presents the changes in the average number of vehicle revenue miles (VRM) over time. Based on the figure below, it can be interpreted that on average, the number of VRM has gradually increased over time.

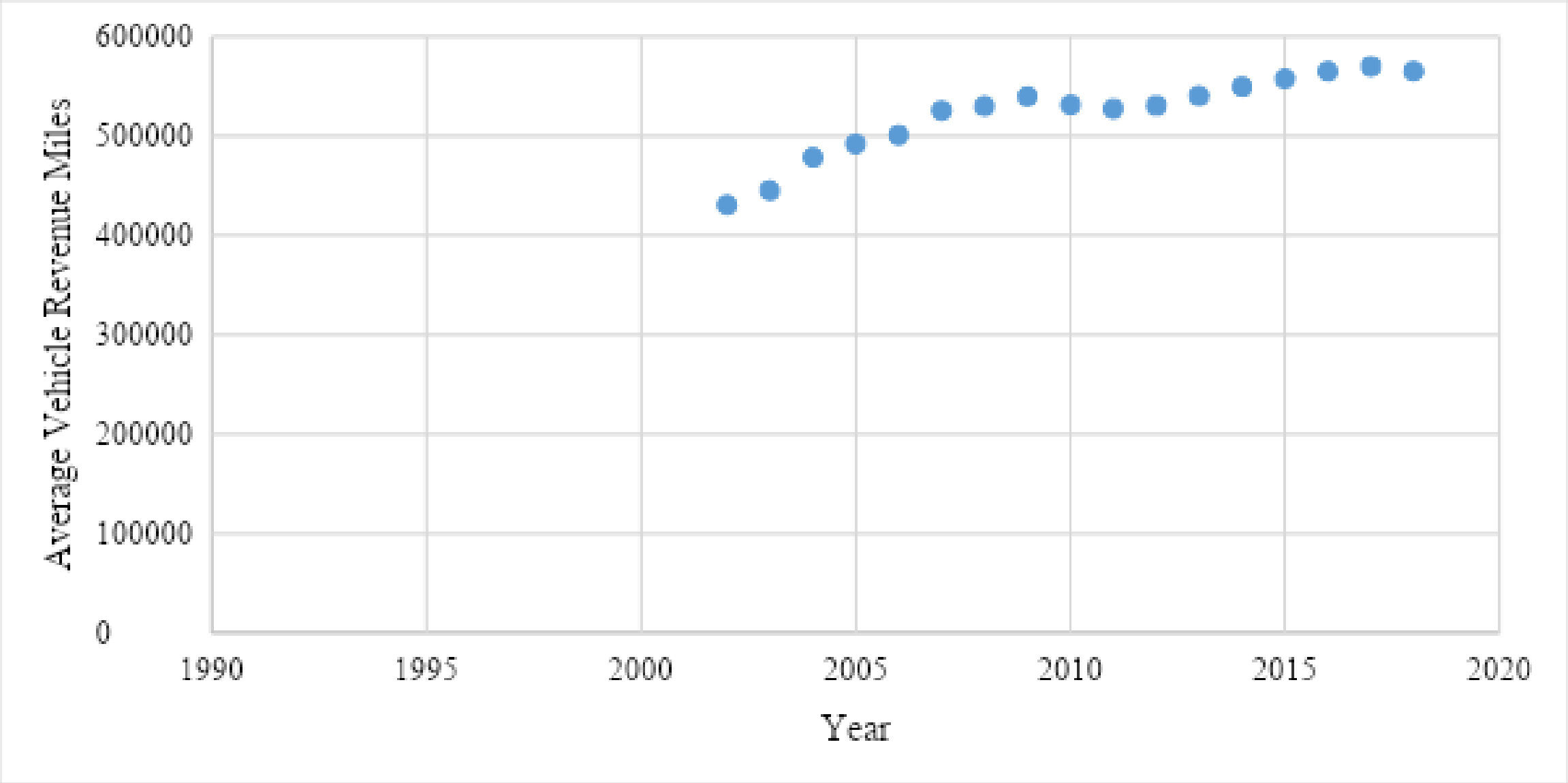

Figure 9: Average Number of Vehicles Revenue Miles over Time

# METHODOLOGY

## Two-Stage Least Squares Regression

For this project, a multiple linear regression modeling procedure, utilizing an ordinary least squares (OLS) estimation technique, is used to identify the factors that influence urban transit ridership demand. It should be noted that the demand for urban transit ridership depends on the presence of transit supply. This simultaneity between transit supply and demand suggests that a 2SLS regression modeling procedure should be conducted. As such, two multiple linear regression models should be developed, one to predict transit supply and a second to predict transit demand, as follows:

$$Y_s = \alpha_s + \beta_s X_s + \varepsilon \#(1)$$

$$Y_D = \alpha_D + \beta_{\hat{Y}_s} \hat{Y}_s + \beta_D X_D + \varepsilon \#(2)$$

where $Y_s$ is the transit supply, $Y_D$ is the transit demand, $\alpha_s$ is the model constant for transit supply, $\alpha_D$ is the model constant for transit demand, $\beta_s$ is the vector of parameter estimates for transit supply, $\beta_D$ is the vector of parameter estimates for transit demand, $X_s$ is the vector of independent variables that are used to explain transit supply, $X_D$ is the vector of independent variables that are used to explain transit demand, $\hat{Y}_s$ is the vector of estimated values for $Y_s$, $\beta_{\hat{Y}_s}$ is the parameter estimate associated with $\hat{Y}_s$, and $\varepsilon$ is the random error term.

It should be noted that a given constant or parameter may be specified in a model yet emerge in the statistical analysis to be statistically equivalent to zero. A student t-test is conducted to evaluate a parameter's statistical significance, and is defined as follows:

$$t = \frac{\hat{\beta}}{s(\hat{\beta})} \#(3)$$

where $s(\hat{\beta})$ is the estimated standard error of $\hat{\beta}$. An estimated constant or variable-coefficient is deemed to be significant if its t-statistic is equal to or greater than the critical t-statistic for a given significance level.

**Goodness-of-Fit Measures**

Various measures can be used when analyzing a multiple linear regression model to determine if the parameters specified are able to adequately predict the values of the dependent variable. These measures are the adjusted R-squared ($R_a^2$), mean absolute error (MAE), and root mean square error (RMSE). The adjusted R-squared, also known as the adjusted coefficient of determination, represents the proportionate reduction of variation from the dependent variable that is associated with the set of explanatory variables used in the model. It may assume a value between zero and one, inclusive, and is expressed as follows:

$$R_a^2 = 1 - \frac{\left(\frac{\sum_{i=1}^{n}\left(y_i - \hat{y}_i\right)^2}{n-K}\right)}{\left(\frac{\sum_{i=1}^{n}\left(y_i - \bar{y}\right)^2}{n-1}\right)} \#(4)$$

Where $y_i$ is the observed response variable, $\hat{y}_i$ is the estimated response variable, $n$ is the number of data entries, $K$ is the number of parameters specified in the model (including the constant term), and $\bar{y}$ is the mean number of the response variable. This measure considers the number of explanatory variables specified in the model and penalizes model specifications with a larger number of explanatory variables. Therefore, the adjusted R-squared can be used as a measure to compare the explanatory power of alternative model specifications.

The MAE is a measure that can be used to evaluate the accuracy of prediction for a specified model. It describes the average magnitude of error for a forecast and can be expressed as follows:

$$MAE = \frac{\sum_{i=1}^{n}\left|y_i - \hat{y}_i\right|}{n} \#(5)$$

However, the MAE does not always provide a clear understanding of the influence that larger errors may have on model forecasts. For this reason, the RMSE should be analyzed. The RMSE gives a higher weight to larger errors and can be compared to the MAE to determine if there are large but infrequent errors in model forecasts. The RMSE can be expressed as follows:

$$RMSE = \sqrt{\frac{\sum_{i=1}^{n}\left(y_i - \hat{y}_i\right)^2}{n}} \#(6)$$

Smaller MAE and RMSE values are indicative of a model that is better specified and with more accurate predictions hence, is to be preferred.

**Selection of Potential Model Variables**

Several variables, as mentioned previously, presented the opportunity to be utilized in model development. Before model development could begin, it was necessary to determine which variables were influential in predicting urban transit supply and demand, respectively. Previous studies (Taylor, Miller, Iseki, and Fink, 2008) utilized vehicle revenue hours (VRH) as the response variable for transit supply. This information was handily available in the NTD files provided, and thus was used as the response variable for the supply model developed in this study. Factors deemed influential to urban transit supply that were handily available in the NTD files included service area density, total average cost per trip, and the average number of vehicles operated in maximum service (VOMS). These variables may then be used to determine a most appropriate model specification to predict transit supply.

The response variable for the demand model should be total unlinked passenger trips (TUPT). Factors deemed influential to urban transit demand that were handily available in the NTD files included the estimated VRH outputs from the supply model, total average fares per trip, and total average trip length. These variables may then be used to determine a most appropriate model specification to predict transit demand.

It should be noted that the scales of many variables are vastly different. As such, log-transformations of variables should be conducted such that appropriate model estimation procedures may be conducted. Furthermore, the log-transformation of variables will allow for easier interpretations of model specifications.

## RESULTS & DISCUSSION OF RESULTS

This section presents the results and discussions of the various models that were developed in this study. The variables that were used in these models are defined in Table 1 below.

Table 1: List of Variables and their Definitions

| Variable | | Definition |
|---|---|---|
| Dependent | VRH | Vehicle revenue hours |
| | TUPT | Total unlinked passenger trip |
| Independent | ACPT | Average cost per trip |
| | SAD | Service area density |
| | AVOMS | Average vehicles operated in maximum service |
| | EVRH | Estimated vehicle revenue hours |
| | AFPT | Total FY average fares per trip |

In this study, several demand and supply models were developed but only the final model specifications are presented in this chapter.

**Supply Model Specification**

In developing the supply model, the dependent variable was VRH and the independent variables were ACPT, SAD, and AVOMS. Table 2 presents the supply model specifications. All estimated coefficients for the independent variables were all positive. This implies that on an average, and an increase in ACPT, SAD, and AVOMS result in an increase in VRH. All estimated coefficients had the right sign and were consistent with previous studies. The goodness of fit (adjusted r-squared) measure for this model was about 0.552 (55.2%). This means that the model is capable of explaining about 55% of the variability in the dataset. The value gotten for the goodness of fit was considerable high since it falls within the range of values of the goodness of fit reported by other studies. All estimated coefficients were found to be significantly different from zero.

Table 2: Parameter Estimates for the Supply Model by Regression Analysis

| **Parameter** | **Estimate** | **t-value** | **t-critical** | **Decision** |
|---|---|---|---|---|
| Intercept | 4.06 | 11.17 | 1.96 | Significant |
| ACPT | 0.22 | 3.62 | 1.65 | Significant |
| SAD | 0.12 | 3.62 | 1.65 | Significant |
| AVOMS | 0.14 | 26.16 | 1.65 | Significant |
| **Adjusted $R^2$** | **0.552** | | | |
| **MAE** | **0.61** | | | |
| **RMSE** | **1.18** | | | |
| **Observations** | **606** | | | |

**Demand Model Specification**

In developing the model, the dependent variable was TUPT and the independent variables were EVRH and AFPT. Table 3 presents the demand model specifications. The estimated coefficient of EVRH was positive. This implies that on an average, and an increase in EVRH will result in an increase in TUPT. Also, the estimated coefficient of AFPT was negative. This implies that on average an increase in AFPT will lead to a decrease in TUPT. All estimated coefficients had the right sign and were consistent with the literature. The goodness of fit (adjusted r-squared) measure for this model was about 0.461 (46.1%). This means that the model is capable of explaining about 46% of the variability in the dataset. The value gotten for the goodness of fit was considerable high since it falls within the range of values of the goodness of fit reported by other studies. All estimated coefficients were found to be significantly different from zero.

Table 3: Parameter Estimates for the Demand Model by Regression Analysis

| Parameter | Estimate | t-value | t-critical | Decision |
| --- | --- | --- | --- | --- |
| Intercept | 5.38 | 13.66 | 1.96 | Significant |
| EVRH | 0.98 | 22.78 | 1.65 | Significant |
| AFPT | -0.13 | -1.99 | 1.65 | Significant |
| **Adjusted $R^2$** | **0.461** | | | |
| **MAE** | **1.04** | | | |
| **RMSE** | **1.37** | | | |
| **Observations** | **606** | | | |

## CONCLUSIONS & RECOMMENDATIONS

In this study, multiple linear regression analysis was used to identify the factors which influence the demand for transit ridership. As the demand for transit ridership is dependent on the supply, a 2SLS regression modeling procedure was used in this study. The following conclusions were made from this study:

- ACPT, SAD, and AVOMS should be used to model urban transit supply;
- EVRH and AFPT should be used to model urban transit demand;
- The supply model has higher explanatory power but is influenced by larger errors, as evidenced by the magnitude of the difference in the MAE and RMSE; and

- The demand model has lower explanatory power compared to the supply model but is not influenced as heavily by larger errors, as evidenced by the magnitude of the difference in the MAE and RMSE.

For further studies and to improve upon the model forecasts, appropriate datasets that contain additional information regarding transit's travel time and socioeconomic characteristics can be useful.

## LIST OF REFERENCES